\documentclass[english,aps,pra,a4paper,twocolumn,showpacs,amsfonts,amsmath,amssymb,amscd,nofootinbib]{revtex4-1}
\usepackage[english]{babel}
\usepackage{graphicx}
\usepackage[dvips]{color}
\usepackage{mathrsfs}
\usepackage{enumerate}
\usepackage{color}

\begin{document}

\def\Journal#1#2#3#4#5#6#7{{#1}, {\it #4} \textbf{#5}, #6 (#2).}
\def\Book#1#2#3#4#5{{#1}, {\it #3} (#4, #5, #2)}
\def\Bookwd#1#2#3#4#5{{#1}, {\it #3} (#4, #5, #2)}

\newcommand{\dd}{\mbox{d}}
\newcommand{\EE}{\mathbb{E}}
\newcommand{\NN}{\mathbb{N}}
\newcommand{\PP}{\mathbb{P}}
\newcommand{\RR}{\mathbb{R}}
\newcommand{\TT}{\mathbb{T}}
\newcommand{\ZZ}{\mathbb{Z}}
\newcommand{\uu}{\mathbf{1}}
\newcommand{\HH}{\mathcal{H}}

\title{Unidirectional quantum walks: evolution and exit times}%
\author{Miquel Montero}
\email[E-mail: ]{miquel.montero@ub.edu}
\affiliation{Departament de F\'{\i}sica Fonamental, Universitat de Barcelona (UB), Mart\'{\i} i Franqu\`es 1, E-08028 Barcelona, Spain}
\date{\today}
\begin{abstract}
In this paper we focus our attention on a particle that follows a unidirectional quantum walk, an alternative version of the nowadays widespread discrete-time quantum walk on a line. Here the walker at each time step can either remain in place or move in a fixed direction, e.g., rightward or upward. While both formulations are essentially equivalent, the present approach leads to consider Discrete Fourier Transforms, which eventually results in obtaining explicit expressions for the wave functions in terms of finite sums, and allows the use of efficient algorithms based on the Fast Fourier Transform. 
The wave functions here obtained govern the probability of finding the particle at any given location, but determine as well the exit-time probability of the walker from a fixed interval, which is also analyzed.
\end{abstract}
\maketitle
\section{Introduction}

Quantum walks~\cite{ADZ93,NK03,JK03,FAJ04}, the quantum mechanical version of the classical random walk |the trajectory of a particle that at each time step moves either leftward or rightward a fixed distance| have attracted the interest of many researchers from heterogeneous areas in recent times: see, e.g.,~\cite{VA12} and references therein. 

We can trace back the origins of quantum walks to quantum computation, the scientific field devoted to build and manage quantum computers, information processing systems whose operation cannot be properly understood without the aid of quantum mechanics. The great potential of such quantum computing devices lies in their capability of running quantum algorithms, algorithms that can be more efficient than those executed by digital computers~\cite{PS97,FG98}.  

The design of new quantum algorithms is not an easy task, as many quantum properties are striking and strongly counterintuitive. A possible approach to this issue is through random walks, since random walks have proved in the past to be a very powerful method for developing algorithms for dealing with a wide range of situations~\cite{MR95}. Among them, the search algorithms deserve special consideration: The left-right random movements of the walker are very well suited to this problem. Therefore, it is not surprising that some of the first applications of quantum walks had this particular purpose~\cite{CFG03,SKW03,AMB10,MNRS11}.

Within this context, and considering the translational invariance of the system, it is quite natural to disregard different alternative formulations for the quantum random walk, as the one we are going to present here: in our version the particle may either move rightward or remain still. This mere change of perspective can encourage, on the one side, the use of computational methods not exploited before and, on the other side, the search for new applications of quantum walks: For instance, thanks to its non-decreasing nature, our process could serve as a {\it quantum subordinator\/}~\cite{A05}. 
 
Subordination replaces each clock tick with time interval that needs a certain stochastic process (the subordinator) to reach or surpass a given point. Once again, hitting-time problems are not new within the framework of bidirectional quantum walks~\cite{BCGJW04,KB06}, including analysis in presence of moving boundaries~\cite{KS11}: The unidirectional process with a fixed threshold maps into the bidirectional one with a target that approaches with a constant velocity, one length unit per time unit. In fact, Ref.~\cite{KS11} considers specifically this particular choice for the velocity, and rates the instance as trivial because the walker eventually reaches the boundary. As we will show, even thus there are striking features to be uncovered.

The paper is organized as follows. In Sec.~\ref{Sec_process} we introduce the process under study, a one-dimensional discrete-time unidirectional quantum walk, and show the connections with previous works. In Sec.~\ref{Sec_wave}  we present explicit formulas for computing the wave functions in the position domain and the corresponding probability mass functions. In this section we also introduce approximate analytic expressions that provide relevant insights into the most noticeable properties.  Section~\ref{Sec_exit} is devoted to the analysis of the exit-time question: We define the problem, present the solution, compare it with its classical counterpart, and finally introduce asymptotic and heuristic approximations. Conclusions are drawn in Sec.~\ref{Sec_conclusion}, where future perspectives are also sketched. We have left for the Appendices the most technical aspects of our mathematical derivations, to prevent any distraction from the main discussion.

\section{The process}
\label{Sec_process}
We begin this paper with a brief review of the general framework of quantum walks. Most of the information contained in this section may be found in (or easily inferred from) standard references in this field~\cite{NK03,JK03,FAJ04,VA12}.  However, since the formulation we adopt here slightly differs from the most used one, we have tried to compose a self-contained text.
 
Let $\HH_P$ be the Hilbert space of discrete particle positions in one-dimension, spanned by the basis $\left\{|\Psi_n\rangle : n \in \{0\}\cup \ZZ^+\right\}$. 
Let $\HH_C$ be the Hilbert space of chirality, or ``coin" states, spanned by the orthonormal basis $\left\{|0\rangle, |1\rangle\right\}$, a qubit. A unidirectional discrete-time, discrete-space quantum walk on the Hilbert space $\HH\equiv\HH_C\otimes \HH_P$ consists of a unitary operator $\hat{U}_C$ acting on the coin state, {\it the throw of the quantum coin\/}, followed by the deterministic updating of the position depending on the qubit value: $\hat{B} \left(|q\rangle\otimes|\Psi_n\rangle \right)= |q\rangle\otimes|\Psi_{n+q}\rangle$.

Explicitly, $\hat{B}$ is a non-decreasing shift operator defined in $\HH$, which takes the following form: 
\begin{eqnarray}
\hat{B}&\equiv& |0\rangle  \langle 0| \otimes \sum_{n=0}^{\infty} |\Psi_n\rangle\langle\Psi_n|+|1\rangle  \langle 1| \otimes \sum_{n=0}^{\infty} |\Psi_{n+1}\rangle\langle\Psi_n|,\nonumber\\
&\equiv&|0\rangle  \langle 0| \otimes \hat{I}_P+|1\rangle  \langle 1| \otimes \hat{S}_P,
\end{eqnarray} 
where $\hat{I}_P$ and $\hat{S}_P$ are the identity operator and the incremental shift operator, respectively, defined in the position space $\HH_P$. The most general expression for the unitary operator $\hat{U}_C$ is
\begin{eqnarray}
\hat{U}_C&\equiv& e^{i \alpha} \cos \varphi |0\rangle  \langle 0| +e^{i \beta} \sin \varphi |0\rangle  \langle 1| \nonumber \\
&+& e^{-i \beta} \sin \varphi  |1\rangle  \langle 0| - e^{-i \alpha} \cos \varphi  |1\rangle  \langle 1|,
\label{U_coin}
\end{eqnarray}
but, as is commonly done, we will choose a fair coin:
\begin{eqnarray}
\hat{H}_C&\equiv& \frac{1}{\sqrt{2}} |0\rangle  \langle 0| +\frac{1}{\sqrt{2}}  |0\rangle  \langle 1| \nonumber \\
&+& \frac{1}{\sqrt{2}}   |1\rangle  \langle 0| -\frac{1}{\sqrt{2}}   |1\rangle  \langle 1|,
\label{H_coin}
\end{eqnarray}
which corresponds to setting $\varphi=\pi/2$, and $\alpha=\beta=0$ in Eq.~\eqref{U_coin}. The unitary operator $\hat{H}_C$ thus defined is called the Hadamard operator, due to its clear connection with the Hadamard transform. 

Based upon the above, the time-evolution operator $\hat{T}$ of the unidirectional quantum walker reads
\begin{equation}
\hat{T}\equiv \hat{B}\left(\hat{H}_C\otimes\hat{I}_P\right).
\label{hat_T}
\end{equation}
When $\hat{T}$ is applied reiteratively on the initial state of the quantum walker, $|\psi\rangle_{0}\equiv|\psi\rangle_{t=0}$, one recovers the state of the system at time $t$, $|\psi\rangle_t$, 
\begin{equation}
|\psi\rangle_t =\left[ \hat{B}\left(\hat{H}_C\otimes\hat{I}_P\right)\right]^{t}|\psi\rangle_{0}.
\end{equation}
In our case, as the time increases in discrete steps, we set the time units so that the variable $t$ is a nonnegative integer quantity, i.e., $t \in \{0\}\cup \ZZ^+$.

We will assume that the initial position of the quantum walker is totally defined, and located at the origin:
\begin{equation}
\hat{M}_0 |\psi\rangle_{0} = |\psi\rangle_{0},
\label{well_located}
\end{equation}
with
\begin{equation}
\hat{M}_n\equiv \left(|0\rangle  \langle 0| +|1\rangle  \langle 1|\right) \otimes |\Psi_{n}\rangle\langle\Psi_n|,
\label{M_def}
\end{equation}
but, by contrast, that the coin state is in a general superposition of the two possible qubit values, that is,
\begin{equation}
|\psi\rangle_{0} =\left(a |0\rangle + b  |1\rangle\right) \otimes |\Psi_0\rangle,
\label{psi_zero_gen}
\end{equation}
where $a$ and $b$ are two complex coefficients such that 
\begin{equation*}
|a|^2+|b|^2=1.
\end{equation*}
 
Before presenting the main results of this work, let us comment how all our expressions can be eventually connected with those corresponding to more conventional version of the discrete-time quantum walk, in which the $|0\rangle$ state in the qubit causes the walker to move leftward. The simplest way is through the following rule of thumb: for any expression valid at time $t$, replace $|\Psi_n\rangle$ with $|\Psi_{2n-t}\rangle$, and extend the position space to include the states $|\Psi_m\rangle$ with $m\in \ZZ^-$, $\HH_E$. 
In other words, one has to apply the time-dependent shift operator $\hat{D}_t$ defined in $\HH_E$,
\begin{equation}
\hat{D}_t\equiv\hat{I}_C\otimes \sum_{n=-\infty}^{\infty} |\Psi_{2n-t}\rangle\langle\Psi_n|,
\end{equation}
on $|\psi\rangle_t$ to recover the bidirectional results at time $t$.

\section{Wave functions and probabilities}
\label{Sec_wave}
We proceed with our work plan by introducing the wave functions $\psi_{0,1}(n,t)$, the two-dimensional projection of the walker state into the position basis:
\begin{eqnarray}
\psi_{0}(n,t)&\equiv& \langle 0|  \otimes  \langle\Psi_n| \psi\rangle_t, \label{Def_Psi_0}\\
\psi_{1}(n,t)&\equiv& \langle 1|  \otimes  \langle\Psi_n| \psi\rangle_t. \label{Def_Psi_1} 
\end{eqnarray}
The evolution operator $\hat{T}$, Eq.~\eqref{hat_T}, 
induces the following set of recurrence equations on the wave-function components:
\begin{eqnarray}
\psi_{0}(n,t)&=&\frac{1}{\sqrt{2}} \psi_{0}(n,t-1)+\frac{1}{\sqrt{2}} \psi_{1}(n,t-1), \label{Rec_0}\\
\psi_{1}(n,t)&=&\frac{1}{\sqrt{2}} \psi_{0}(n-1,t-1)-\frac{1}{\sqrt{2}} \psi_{1}(n-1,t-1),\nonumber \label{Rec_1} \\
\end{eqnarray}
which are to be solved under the assumption that the walker is initially at $n=0$, that is, $\psi_{0}(n,0)=a\, \delta_{n,0}$, $\psi_{1}(n,0)=b\, \delta_{n,0}$, where $\delta_{n,m}$ is the Kronecker delta.

In Appendix~\ref{App_Exact} we show how the answer to the posed problem reads 
\begin{eqnarray}
\psi_{0}(n,t)&=&\frac{a}{N}\Bigg\{\frac{1+(-1)^t}{2}+\frac{1-(-1)^t}{2\sqrt{2}}\nonumber \\
&+&\sum_{r=1}^{N-1}  \frac{\cos\left[\pi (2n-t)r/N+ \omega_{r/N} t\right]}{2-\sqrt{2} \cos[\omega_{r/N} -\pi r /N]}\Bigg\}\nonumber \\
&+&\frac{b}{N}\Bigg\{\frac{1-(-1)^t}{2\sqrt{2}}\nonumber \\
&+&\sum_{r=1}^{N-1}  \frac{\sqrt{2} \cos \omega_{r/N} -\cos \frac{\pi r}{N}}{2-\sqrt{2} \cos[\omega_{r/N} -\pi r /N]}\nonumber \\
&\times& \cos\left[\pi (2n-t+1)r/N+ \omega_{r/N} t\right]\Bigg\},
\label{Sol_Psi0}
\end{eqnarray}
and
\begin{eqnarray}
\psi_{1}(n,t)&=&\frac{a}{N} \Bigg\{\frac{1-(-1)^t}{2\sqrt{2}}\nonumber \\
&+&\sum_{r=1}^{N-1}  \frac{\sqrt{2} \cos \omega_{r/N} -\cos \frac{\pi r}{N}}{2-\sqrt{2} \cos[\omega_{r/N} -\pi r /N]}\nonumber \\
&\times& \cos\left[\pi (2n-t-1)r/N+ \omega_{r/N} t\right]\Bigg\},\nonumber \\
&+&\frac{b}{N}\Bigg\{\frac{1+(-1)^t}{2}-\frac{1-(-1)^t}{2\sqrt{2}}\nonumber \\
&+&\sum_{r=1}^{N-1}  \frac{\left(\sqrt{2} \cos \omega_{r/N} -\cos \frac{\pi r}{N}\right)^2}{2-\sqrt{2} \cos[\omega_{r/N} -\pi r /N]}\nonumber \\
&\times& \cos\left[\pi (2n-t)r/N+ \omega_{r/N} t\right]\Bigg\},
\label{Sol_Psi1}
\end{eqnarray}
where here $n\in\{0,\ldots,t\}$, $N$ is any natural number greater than $t$, and $\omega_{r/N}$, is the only solution that 
\begin{equation}
\omega_{r/N} =\arcsin\left(\frac{1}{\sqrt{2}}\sin  \frac{\pi r}{N}\right)
\label{omega_r}
\end{equation}
has in the $[0,\pi/4]$ range. Since the final outcome of Eqs.~\eqref{Sol_Psi0} and~\eqref{Sol_Psi1} does not depend on the particular value of $N$, as long as $N>t$, two natural choices arise: $N=t+1$ and $N=2^k$, with $k$ the smallest integer for which it holds $t<2^k$. In Appendix~\ref{App_Exact} we discuss the convenience of the second option, in particular for large values of $t$, because in this case one can benefit from the power of the Fast Fourier Transform algorithm. By contrast, since in every illustrative instance we are going to consider along this paper the value of $t$ is moderately small, we choose $N=t+1$ in practice.

\begin{figure}[htbp]
{
\includegraphics[width=1.0\columnwidth,keepaspectratio=true]{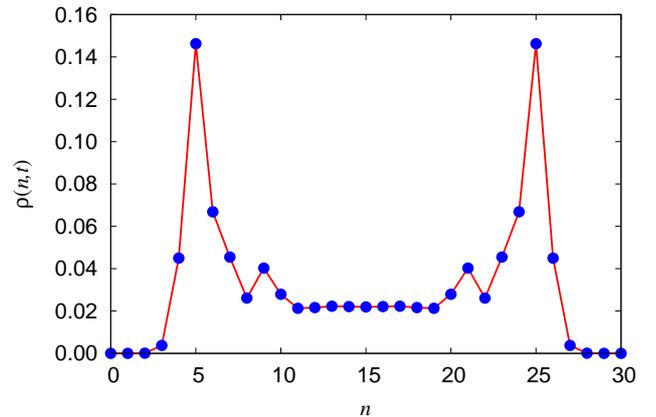}
}
\caption{(Color online) 
Probability mass function of the process for $t=30$ time steps.  The (red) solid line connects the points obtained by direct application of the evolution operator on the initial state. The (blue) circles were computed by means of Eqs.~\eqref{Sol_Psi0} and~\eqref{Sol_Psi1}.} 
\label{Fig_sample}
\end{figure}

In Fig.~\ref{Fig_sample} we present evidence in support of the soundness of the solution shown in Eqs.~\eqref{Sol_Psi0} and~\eqref{Sol_Psi1}. In there, as in the rest of the forthcoming examples, we have considered the following initial state:
\begin{equation}
|\psi\rangle_{t=0} =\frac{1}{\sqrt{2}}\left[ |0\rangle + i  |1\rangle\right] \otimes |\Psi_0\rangle,
\label{psi_zero}
\end{equation}
we have subsequently computed $|\psi\rangle_t$ by systematic application of the translation operator $\hat{T}$, and evaluated the probability that the walker is at any given position, $\rho(n,t)$, the Probability Mass Function (PMF) of the process,
\begin{equation}
\rho(n,t)\equiv  \langle\psi|\hat{M}_n|\psi\rangle_{t} .
\label{rho_def_bracket}
\end{equation}
The result is in excellent agreement with the one obtained through the numerical evaluation of  Eqs.~\eqref{Sol_Psi0} and~\eqref{Sol_Psi1}, for $a=1/\sqrt{2}$ and $b=i/\sqrt{2}$, with
\begin{equation}
\rho(n,t)=\left|\psi_0(n,t)\right|^2+\left|\psi_1(n,t)\right|^2.
\label{rho_def_mods}
\end{equation}

The intricate nature of the formulas shown in Eqs.~\eqref{Sol_Psi0} and~\eqref{Sol_Psi1} motivates the search for simpler expressions, even at the cost of obtaining mere approximations. To this end, it is very convenient to consider the limit $t\gg 1$, $n\gg 1$, but by keeping $\nu\equiv n/t$ finite. In Appendix~\ref{App_App} we show how, under the previous assumptions, $\rho(n,t)$ can be approximated by $\bar{\rho}(n,t)$,
\begin{eqnarray}
\bar{\rho}(n,t)&\equiv&\frac{1}{t} \frac{1}{2\pi \nu(1-\nu)\sqrt{1-2(1-2\nu)^2}} \nonumber \\
&\times&\left\{1+2(1-2\nu)^2\sin\left[2\phi_0(\nu) t\right]\right\},
\label{Prob_App_Main}
\end{eqnarray}
where
\begin{eqnarray}
\phi_0(\nu) &\equiv& (2\nu-1) \arcsin\left(\sqrt{\frac{1-2(1-2\nu)^2}{4\nu(1-\nu)}}\right)\nonumber\\
&+&\arcsin\left(\sqrt{\frac{1-2(1-2\nu)^2}{8\nu(1-\nu)}}\right),
\end{eqnarray}
as long as the expressions under the root signs remain positive, that if, for
\begin{equation}
\frac{1}{2}\left(1-\frac{1}{\sqrt{2}}\right)<\nu<\frac{1}{2}\left(1+\frac{1}{\sqrt{2}}\right).
\label{nu_limits}
\end{equation} 

Let us analyze the structure of Eq.~\eqref{Prob_App_Main}. The presence of a sinusoidal term in $\bar{\rho}(n,t)$ leads to the natural definition of $\bar{\rho}_{\rm max}(n,t)$, 
\begin{eqnarray}
\bar{\rho}_{\rm max}(n,t)&\equiv&\frac{1}{t} \frac{1+2(1-2\nu)^2}{2\pi \nu(1-\nu)\sqrt{1-2(1-2\nu)^2}},
\label{Prob_App_Sup_Main}
\end{eqnarray}
and $\bar{\rho}_{\rm min}(n,t)$,
\begin{eqnarray}
\bar{\rho}_{\rm min}(n,t)&\equiv&\frac{1}{t} \frac{1-2(1-2\nu)^2}{2\pi \nu(1-\nu)\sqrt{1-2(1-2\nu)^2}} \nonumber \\
&=&\frac{1}{t} \frac{\sqrt{1-2(1-2\nu)^2}}{2\pi \nu(1-\nu)},
\label{Prob_App_Inf_Main}
\end{eqnarray}
in such a way $\bar{\rho}_{\rm max}(n,t)\leq \bar{\rho}(n,t) \leq \bar{\rho}_{\rm min}(n,t)$. 

Unlike $\bar{\rho}(n,t)$ itself, $\bar{\rho}_{\rm max}(n,t)$ and $\bar{\rho}_{\rm min}(n,t)$ have not been previously reported in the literature,~\footnote{The key point to understand this fact can be found in the functional form of earlier expressions of $\bar{\rho}(n,t)$ which, although ultimately equivalent to the present one, failed to concentrate all the oscillatory behavior in a single term.} and clarify the origin of some of the most distinctive traits of the position PMF. In particular, they are 
well suited to quantify the ``quasi-uniform behavior''~\cite{VA12} of $\rho(n,t)$, its apparent lack of dependence on $n$, for $n\sim t/2$, that was already present in Fig.~\ref{Fig_sample}. If one expands $\bar{\rho}_{\rm max}(n,t)$ and $\bar{\rho}_{\rm min}(n,t)$, for a fixed $t$, around $n=t/2$, one finds 
\begin{eqnarray}
\bar{\rho}_{\rm max}(n,t)&\sim&\frac{2}{\pi t} \left[1+16\, \varepsilon^2\right],\\
\bar{\rho}_{\rm min}(n,t)&\sim&\frac{2}{\pi t} \left[1-8\,{\varepsilon}^4\right],
\end{eqnarray}
with 
$\varepsilon\equiv \nu-1/2$. Since we have ${\varepsilon}^4<1/64$, cf. Eq.~\eqref{nu_limits}, this means that $\bar{\rho}_{\rm min}(n,t)\sim \frac{2}{\pi t}$ is a good approximation for the entire region.   

\begin{figure}[htbp]
{
\includegraphics[width=1.0\columnwidth,keepaspectratio=true]{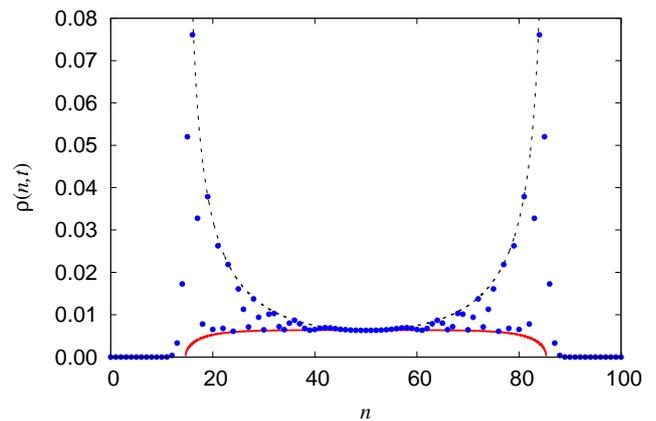}
}
\caption{(Color online) Probability mass function on the process for $t=100$ time steps.  The (blue) circles were obtained by direct evaluation of $\rho(n,t)$,  Eq.~\eqref{rho_def_mods}. 
The (black) dashed line corresponds to $\bar{\rho}_{\rm max}(n,t)$, Eq.~\eqref{Prob_App_Sup_Main}, whereas the (red) solid line indicates the lower approximate value $\bar{\rho}_{\rm min}(n,t)$, Eq.~\eqref{Prob_App_Inf_Main}.} 
\label{Fig_min_max}
\end{figure}

In Fig.~\eqref{Fig_min_max} we have evolved the initial state in Eq.~\eqref{psi_zero} up to time $t=100$, and represented $\rho(n,t)$. Observe how all the points almost perfectly accommodate within the limits marked by $\bar{\rho}_{\rm max}(n,t)$ and $\bar{\rho}_{\rm min}(n,t)$, which is nearly flat along the domain where this function is well defined.

\section{Exit-time probabilities}
\label{Sec_exit}
The expressions in Eqs.~\eqref{Sol_Psi0} and~\eqref{Sol_Psi1} completely determine the evolution of the system, not just the spatial PMF, and therefore we can use them to solve a different but related problem: the computation of the exit-time probability.

The concepts of {\it exit time\/}, $\mathcal{T}_\mathcal{D}$, and exit-time probability,
\begin{equation}
\mathcal{P}_{\mathcal{D}}(t)\equiv \Pr\left\{\mathcal{T}_\mathcal{D}=t\right\},
\end{equation}   
are clear within the context of stochastic processes: they are related to the random instant at which the process leaves a given domain $\mathcal{D}$ for the first time. The quantum nature of the walker forces us to consider a more accurate definition, because one cannot know if the walker has left the region unless some measure is performed. Since the measuring act modifies the state of the system, the way in which we determine if the process remains within the region will affect the very exit time. 

In this case, we consider that the domain is set equal to the interval $\mathcal{D}\equiv[0,n_0)$, with $n_0\geq 1$, and all the probability is initially concentrated at the origin,  Eq.~\eqref{psi_zero}. The only way our quantum walker can escape form the interval is through $n_0$. Note that $\psi_0(n_0,t)=0$ for $t\leq n_0$, and  $\psi_1(n_0,t)=0$ for $t< n_0$. Therefore, the first chance for the system to leave the region is when $t=n_0$, because $\psi_1(n_0,n_0)\neq 0$. Let us assume that at this time we measure if the walker is at $n=n_0$. If the answer is ``yes", 
the exit time is simply $\mathcal{T}_{[0,n_0)}=n_0$, and the corresponding exit-time probability reads $\mathcal{P}_{[0,n_0)}(n_0)=\left|\psi_1(n_0,n_0)\right|^2$. If the answer is ``no", the wave function is filtered, the $\psi_1(n_0,n_0)$ contribution is removed from the wave function, $\psi^*_1(n_0,n_0)=0$, $\psi^*_0(n_0,n_0)=\psi_0(n_0,n_0)=0$, and
\begin{equation}
\psi^*_{0,1}(n,n_0)\equiv\frac{\psi_{0,1}(n,n_0)}{\sqrt{1-\left|\psi_1(n_0,n_0)\right|^2}},
\end{equation}
for $n<n_0$. At the next time step we will have 
 \begin{eqnarray*}
\psi^*_0(n,n_0+1)&=&0,\\
\psi^*_{1}(n,n_0+1)&=&\frac{\psi_{1}(n,n_0+1)}{\sqrt{1-\left|\psi_1(n_0,n_0)\right|^2}},
\end{eqnarray*}
cf. Eqs.~\eqref{Rec_0} and ~\eqref{Rec_1}. Following the same reasoning as above, the {\it conditional\/} exit-time probability is equal to $\left|\psi^*_1(n_0,n_0+1)\right|^2$, and  thus the  exit-time probability reduces to
\begin{equation}
\mathcal{P}_{[0,n_0)}(n_0+1)=\left|\psi_1(n_0,n_0+1)\right|^2.
\end{equation}
Now we can simply iterate the argument, and conclude that 
\begin{equation}
\mathcal{P}_{[0,n_0)}(t)=\left|\psi_1(n_0,t)\right|^2,
\end{equation}
for $t\geq n_0$.

\begin{figure}[htbp]
{
\includegraphics[width=1.0\columnwidth,keepaspectratio=true]{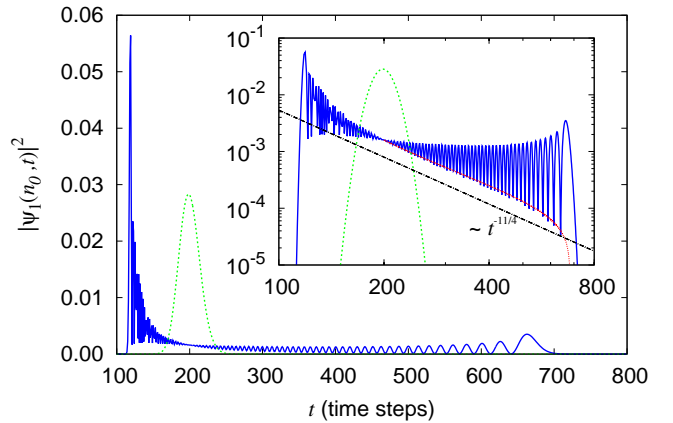}
}
\caption{(Color online) Exit times for $n_0=100$. Main plot: The exit-time probability of the quantum walker, the (blue) continuous line, is compared with the classical result, the (green) dashed line. Inset: The same probabilities in a double logarithmic scale. The (red) dotted line is the approximate lower bound, Eq.~\eqref{Exit_Prob_App_Inf_Main}, whereas the (black) dot-dashed line serves as a guide for the eye, cf. Eq.~\eqref{Exit_Prob_Heur_Inf_Main}.} 
\label{Fig_exit_time}
\end{figure}

In Fig.~\ref{Fig_exit_time} we present the exit-time probability when $n_0=100$. As it can be observed, the probability $\mathcal{P}_{[0,n_0)}(t)$ noticeably differs from its classical counterpart,   
\begin{equation}
\mathcal{P}_{[0,n_0)}^{\rm clas}(t)=\binom{t-1}{t-n_0} p^{n_0}(1-p)^{t-n_0},
\end{equation}
where $p$ is probability that the walker changes its location, $p=1/2$ here, and $t\geq n_0\geq 1$. The classical exit-time probability is bell-shaped around $t=2 n_0$, whereas the quantum probability attains its maximum short after $t=n_0$, a reminiscence of the functional form of $\rho(n,t)$. In fact, the time $t=2 n_0$ marks the instant after which the behavior of $\mathcal{P}_{[0,n_0)}(t)$ changes qualitatively. Equation~\eqref{Psi_1_App} in Appendix~\ref{App_App} shows the origin of this point of inflexion. Note that $t=2 n_0$ corresponds to $\nu=1/2$, and the two cosine terms in Eq.~\eqref{Psi_1_App} have exactly the same weight, whereas for $\nu\neq 1/2$ the global behavior is dominated by either one or the other. Also in Appendix~\ref{App_App} we can find that for $t>2 n_0$ we have the following approximate lower bound for $\mathcal{P}_{[0,n_0)}(t)$, 
\begin{equation}
\mathcal{P}_{[0,n_0)}(t)\gtrsim \frac{1}{4 \pi} \frac{\sqrt{8 n_0 (t-n_0)-t^2}}{(t-n_0)^2},
\label{Exit_Prob_App_Inf_Main} 
\end{equation}
an expression that captures the decay rate of the exit-time probability, as can be checked in the inset of Fig.~\ref{Fig_exit_time}. As a final curiosity, this decay rate seems in good agreement with the heuristic expression
\begin{equation}
\mathcal{P}_{[0,n_0)}(t)\sim \frac{1}{2 \pi n_0} \left(\frac{2 n_0}{t}\right)^{\frac{11}{4}},
\label{Exit_Prob_Heur_Inf_Main} 
\end{equation}
a fact without a clear explanation.

\section{Conclusion}
\label{Sec_conclusion}

In this paper we have analyzed the unidirectional quantum walk, an alternative formulation of the discrete-time, discrete-space, quantum walk on a line in which the walker can either remain in place or proceed in a fixed direction but never move backward. Here lies the main difference with respect to the most typical setup where the particle can move in either direction. The translational invariance of the problem makes both formalisms essentially equivalent and every formula or property can be easily rephrased, a fact that adds value to our results. 

The most prominent of these results is the derivation of exact algebraic expressions for the wave functions that govern the probability of finding the particle at any given location, the probability mass function. These formulas are originally based on the Discrete Fourier Transform which allows the use of efficient algorithms based on the Fast Fourier Transform to evaluate them. 

A second interesting result, related to the previous one, is the obtaining of two approximate functions that limit the range of variation of the probability mass function. These functions contain the clue to the understanding of the ``quasi-uniform behavior" of this probability.
   
The third outstanding result is the possible use of unidirectional quantum walks as stochastic subordinators. The study of the probability of the exit time of the process out of a fixed interval indicates the presence of a transient period in which the exit probability decays algebraically, with an effective rational exponent of uncertain origin. Clearly, this phenomenon deserves further attention but is left for a future work. 

\begin{acknowledgments}
The author acknowledges partial support from the former Spanish Ministerio de Ciencia e Innovaci\'on under Contract No. FIS2009-09689, and from Generalitat de Catalunya, Contract No. 2009SGR417.
\end{acknowledgments}

\appendix
\section{General solution}
\label{App_Exact}
In this Appendix we provide further details on the derivation of the explicit expressions for the two components of the wave functions given in the main text, Eqs.~\eqref{Sol_Psi0} and~\eqref{Sol_Psi1}, starting from the recurrence formulas~\eqref{Rec_0} and~\eqref{Rec_1}. The approach that follows, like the one taken in previous references~\cite{VA12} is based on the Fourier analysis. The main difference lies in an apparently subtle change: we have decided to use the Discrete Fourier Transform (DFT) instead of the Discrete-Time Fourier Transform (DTFT). This choice is inspired (but not forced) by the fact that within our formulation $\psi_{0}(n,t)$ and $\psi_{1}(n,t)$ only take values different from zero when $n$ is a nonnegative integer.   

Be $f(n)$ a complex function, $n \in \{0,\ldots, N-1\}$, and denote by $\tilde{f}(r)$ its DFT,
\begin{equation}
\tilde{f}(r)\equiv \sum_{n=0}^{N-1} f(n) e^{i 2 \pi r n/N},
\label{DFT}
\end{equation} 
for $r\in \{0,\ldots, N-1\}$. Then, it is well known that one can recover $f(n)$ from $\tilde{f}(r)$ by means of the inverse DFT expression
\begin{equation}
f(n)\equiv \frac{1}{N}\sum_{r=0}^{N-1} \tilde{f}(r) e^{-i 2 \pi r n/N}.
\label{IDFT}
\end{equation}
The inversion formula contains the most perceptible discrepancy between DFTs and DTFTs: in the latter, $r$ is a continuous index and the inversion procedure involves the computation of definite integrals instead of finite sums. The truth is that, whereas DTFTs assume that the original function is a periodical magnitude sampled at regular intervals, DFTs impose no restriction on $f(n)$, not even that $f(n)=0$ when $n<0$ or $n\geq N$: all the information contained in the $N$ complex numbers $f(n)$ is directly mapped into the $N$ complex quantities $\tilde{f}(r)$.     
 
Thus, our next step is to decide a suitable value for $N$. In this case, since $\psi_{0,1}(n,t)=0$ for $n\geq t+1$, we could simply set $N=t+1$. However, this is not a very convenient choice if we want to transform our set of two recurrence equations in the position domain into a set of algebraic equations in the Fourier domain: These recurrence formulas involve, not only different locations but different instants of time, cf. Eqs.~\eqref{Rec_0} and~\eqref{Rec_1}, and linking $N$ and $t$ prevents us from achieving our goal. 

To avoid that, let us introduce the auxiliary time horizon $T$, $T\geq 0$, set $N\equiv T+1$, and 
consider the following definition for the DFT of $\psi_{0,1}(n,t)$, valid for any $t$, $t\in\{0,\ldots, T\}$,~\footnote{For notational convenience, $N$ and $T$ may alternate or even coexist in expressions appearing along this Appendix.}
\begin{equation}
\tilde{\psi}_{0,1}(r,t;T)\equiv \sum_{n=0}^{N-1} \psi_{0,1}(n,t) e^{i 2 \pi r n/N}.
\label{DFT_Psi}
\end{equation}
Note that, while $\tilde{\psi}_{0,1}(r,t;T)$ is an explicit function of $T$ |that is, for a fixed value of $r$ and a fixed value of $t$, different choices of $T$ lead to different values for $\tilde{\psi}_{0,1}(r,t;T)$|, the final result of applying the corresponding inversion formula,
\begin{equation}
\psi_{0,1}(n,t)\equiv \frac{1}{N}\sum_{r=0}^{N-1} \tilde{\psi}_{0,1}(r,t;T) e^{-i 2 \pi r n/N},
\label{IDFT_Psi}
\end{equation}
does not depend on $T$, for a fixed choice of $n$ and $t$, as long as one restricts 
these two variables to be in the set $\{0,\ldots,T\}$. Obviously, given $t$, if one evaluates Eq.~\eqref{IDFT_Psi} for $n=t+1,\ldots,N-1$, one will obtain $\psi_{0,1}(n,t)=0$ identically. Therefore, in principle, there is no reason to compute Eq.~\eqref{IDFT_Psi} out of the range $n=0,\ldots,t$. We will return to this issue later on.

At this point we can move Eqs.~\eqref{Rec_0} and~\eqref{Rec_1} into the Fourier domain:
\begin{eqnarray}
\tilde{\psi}_{0}(r,t;T)&=&\frac{1}{\sqrt{2}} \tilde{\psi}_{0}(r,t-1;T)\nonumber \\
&+&\frac{1}{\sqrt{2}} \tilde{\psi}_1(r,t-1;T), \label{FRec_0}\\
\tilde{\psi}(r,t;T)&=&\frac{e^{i 2 \pi r/N}}{\sqrt{2}} \tilde{\psi}_{0}(r,t-1;T)\nonumber \\
&+&\frac{e^{i 2 \pi r/N}}{\sqrt{2}} \tilde{\psi}_{1}(r,t-1;T).\label{FRec_1}
\end{eqnarray}
The initial values for $\tilde{\psi}_{0,1}(r,t;T)$ are $\tilde{\psi}_{0}(r,0;T)=a$, $\tilde{\psi}_{1}(r,0;T)=b$, for $r \in\{0,\ldots,N-1\}$. The resolution of Eqs.~\eqref{FRec_0} and~\eqref{FRec_1} can be tackled through standard matrix techniques, thus resulting in
\begin{eqnarray}
\tilde{\psi}_{0}(r,t;T)&=&\frac{(\lambda_+)^t}{1+|1-\sqrt{2} \lambda_+|^2} \left[a + (\sqrt{2} \lambda_+^{-1}-1)b\right]\nonumber \\
&+&\frac{(\lambda_-)^t}{1+|1-\sqrt{2} \lambda_-|^2} \left[a + (\sqrt{2} \lambda_-^{-1}-1)b\right],\nonumber \\ \label{Sol_Fpsi0}
\end{eqnarray}
and
\begin{eqnarray}
\tilde{\psi}_{1}(r,t;T)&=&\frac{(\lambda_+)^t(\sqrt{2} \lambda_+-1)}{1+|1-\sqrt{2} \lambda_+|^2} \left[ a + (\sqrt{2} \lambda_+^{-1}-1)b\right]\nonumber \\
&+&\frac{(\lambda_-)^t(\sqrt{2} \lambda_--1)}{1+|1-\sqrt{2} \lambda_-|^2} \left[a + (\sqrt{2} \lambda_-^{-1}-1)b\right],\nonumber \\ \label{Sol_Fpsi1}
\end{eqnarray}
with the dependence on $r$ and $T$ hidden in $\lambda_+$and $\lambda_-$, 
\begin{eqnarray}
\lambda_+&\equiv&e^{-i(\omega_{r/N} -\pi r /N)},\\
\lambda_-&\equiv&-e^{i(\omega_{r/N} +\pi r /N)},
\end{eqnarray}
and where $\omega_{r/N}$ is an angle that, given $r$ and $N$, satisfies
\begin{equation}
\sin \omega_{r/N} =\frac{1}{\sqrt{2}}\sin  \frac{\pi r}{N}.
\label{App_omega_r}
\end{equation}
Note that, since $r\in\{0\ldots,N-1\}$, we have
\begin{equation*}
0\leq \sin \omega_{r/N} \leq \frac{1}{\sqrt{2}}, 
\end{equation*}
so, to prevent any uncertainty, we consider that $\omega_{r/N}$ is the only solution that Eq.~\eqref{App_omega_r} has in $[0,\pi/4]$. 

Now, we can simply introduce the expressions of $\tilde{\psi}_{0,1}(r,t;T)$ in~\eqref{Sol_Fpsi0} and~\eqref{Sol_Fpsi1} into Eq.~\eqref{IDFT_Psi} and recover $\psi_{0,1}(n,t)$ after the computation of a finite sum.  To manage the complexity of these expressions we analyze the particular case $a=1$ and $b=0$  in the first place: 
\begin{eqnarray}
\psi_{0}(n,t)&=&\frac{1}{N}\sum_{r=0}^{N-1}  \frac{e^{-i\left[\pi (2n-t)r/N+ \omega_{r/N} t\right]}}{1+|1-\sqrt{2} e^{-i(\omega_{r/N} -\pi r /N)}|^2}\nonumber \\
&+&\frac{(-1)^t}{N}\sum_{r=0}^{N-1}  \frac{e^{i\left[-\pi (2n-t)r/N+ \omega_{r/N} t\right]}}{1+|1+\sqrt{2} e^{i(\omega_{r/N} +\pi r /N)}|^2}.\nonumber \\ 
\end{eqnarray}
We begin by detaching the $r=0$ term from the summations above
\begin{eqnarray*}
\psi_{0}(n,t)&=&\frac{1}{N}\frac{1}{4-2\sqrt{2}}+\frac{(-1)^t}{N} \frac{1}{4+2\sqrt{2}}\nonumber\\
&+&\frac{1}{N}\sum_{r=1}^{N-1}  \frac{e^{-i\left[\pi (2n-t)r/N+ \omega_{r/N} t\right]}}{1+|1-\sqrt{2} e^{-i(\omega_{r/N} -\pi r /N)}|^2}\nonumber \\
&+&\frac{(-1)^t}{N}\sum_{r=1}^{N-1}  \frac{e^{i\left[-\pi (2n-t)r/N+ \omega_{r/N} t\right]}}{1+|1+\sqrt{2} e^{i(\omega_{r/N} +\pi r /N)}|^2}.\nonumber \\ 
\end{eqnarray*}
Now, we can define $s\equiv N-r$ in the last sum and rearrange the whole expression to finally obtain
\begin{eqnarray}
\psi_{0}(n,t)&=&\frac{1}{N}\left(\frac{1+(-1)^t}{2}+\frac{1-(-1)^t}{2\sqrt{2}}\right)\nonumber \\
&+&\frac{1}{N}\sum_{r=1}^{N-1}  \frac{\cos\left[\pi (2n-t)r/N+ \omega_{r/N} t\right]}{2-\sqrt{2} \cos(\omega_{r/N} -\pi r /N)},\nonumber \\
\end{eqnarray}
where the fact that $\omega_{s/N}=\omega_{r/N}$ has been taken into account. A similar procedure give us 
\begin{eqnarray}
\psi_{1}(n,t)&=&\frac{1-(-1)^t}{N}\frac{1}{2\sqrt{2}}\nonumber \\
&+&\frac{1}{N}\sum_{r=1}^{N-1}  \frac{\sqrt{2} \cos \omega_{r/N} -\cos \frac{\pi r}{N}}{2-\sqrt{2} \cos(\omega_{r/N} -\pi r /N)}\nonumber \\
&\times& \cos\left[\pi (2n-t-1)r/N+ \omega_{r/N} t\right],\nonumber \\
\end{eqnarray}
where we have used that
\begin{eqnarray*}
\sqrt{2} \lambda_{+}-1&=&\left(\sqrt{2} \cos \omega_{r/N} -\cos \frac{\pi r}{N}\right) e^{i \pi r /N},\\
\sqrt{2} \lambda_{-}-1&=&\left(\sqrt{2} \cos \omega_{s/N} -\cos \frac{\pi s}{N}\right) e^{-i \pi s /N},\\
\end{eqnarray*}
with the same definition for $s$ as before, $s=N-r$. Analogously, when $a=0$ and $b=1$ we have
\begin{eqnarray}
\psi_{0}(n,t)&=&\frac{1-(-1)^t}{N}\frac{1}{2\sqrt{2}}\nonumber \\
&+&\frac{1}{N}\sum_{r=1}^{N-1}  \frac{\sqrt{2} \cos \omega_{r/N} -\cos \frac{\pi r}{N}}{2-\sqrt{2} \cos(\omega_{r/N} -\pi r /N)}\nonumber \\
&\times& \cos\left[\pi (2n-t+1)r/N+ \omega_{r/N} t\right],\nonumber \\
\end{eqnarray}
and
\begin{eqnarray}
\psi_{1}(n,t)&=&\frac{1}{N}\left(\frac{1+(-1)^t}{2}-\frac{1-(-1)^t}{2\sqrt{2}}\right)\nonumber \\
&+&\frac{1}{N}\sum_{r=1}^{N-1}  \frac{\left(\sqrt{2} \cos \omega_{r/N} -\cos \frac{\pi r}{N}\right)^2}{2-\sqrt{2} \cos(\omega_{r/N} -\pi r /N)}\nonumber \\
&\times& \cos\left[\pi (2n-t)r/N+ \omega_{r/N} t\right].\nonumber \\
\end{eqnarray}
We can recover the general solution,  Eqs.~\eqref{Sol_Psi0} and~\eqref{Sol_Psi1}, through the superposition of these two cases.  

A final remark on the role that plays $N$ in the computational complexity of the results in Eqs.~\eqref{Sol_Psi0} and~\eqref{Sol_Psi1}. For $N$ and $t$ fixed, the number of complex operations one needs to obtain each wave function for every value of $n$, $n\in\{0,\ldots,t\}$, is roughly $\mathcal{O}(t\times N)$. Therefore, one can easily reduce this quantity up to $\mathcal{O}(t^2)$ by setting $N=t-1$. However, if one chooses $N$ such that $N=2^k$, $k\in\NN$,  the whole solution may be recovered by means of the Fast Fourier Transform (FFT) algorithm. Since the computational complexity of this method is just $\mathcal{O}(k \times 2^k)$, it is always worth considering the FFT approach for large values of $t$.

\section{Approximate expressions}
\label{App_App}
In this Appendix we obtain alternative equations for $\psi_{0,1}(n,t)$, expressions that are more compact and readable than Eqs.~\eqref{Sol_Psi0} and~\eqref{Sol_Psi1}, although approximate. 

A close analysis of the inner structure of the four pieces that conform Eqs.~\eqref{Sol_Psi0} and~\eqref{Sol_Psi1} shows us that we must repeatedly analyze functions like $h(n,t)$, 
\begin{eqnarray}
h(n,t)&\equiv&\frac{\Xi(t)}{N}\nonumber\\&+& \frac{1}{N}\sum_{r=1}^{N-1} g(r/N) \cos\left[\theta(n,r,t;T)+\epsilon \pi r/N\right],\nonumber\\
\end{eqnarray}
where 
\begin{equation}
\theta(n,r,t;T)\equiv \pi (2n-t)r/N+ \omega_{r/N} t,
\end{equation}
and $\epsilon\in\{-1,0,1\}$. In every case $g(\cdot)$ is a smooth function, and therefore the behavior of the cosine terms does determine the overall result of the sum. Due to the presence of $\omega_{r/N}$ within $\theta(n,r,t;T)$, the argument of these cosine functions does not change linearly with $r$ but exhibits a maximum, and then the use of a tailored version of the method of the stationary phase is the most indicated in this case~\cite{CH46,VA12}: Only those terms for which $\theta(n,r,t;T)$ attains its maximum are relevant, whereas the rest of them are negligible.   

To this end, let us firstly define $u\equiv r/N$, and $\nu\equiv n/t$, in terms of which we can rewrite $\theta(n,r,t;T)$,
\begin{equation}
\theta(\nu t,u (T-1) ,t;T)=\phi(\nu, u)t,
\end{equation}
with
\begin{equation}
\phi(\nu,u)\equiv \pi (2\nu-1) u+ \omega_u.
\end{equation}
Our next step is to consider function $h(n,t)$ in the continuum limit, $N\to \infty$, 
\begin{eqnarray}
h(n,t)&\sim&\int_0^{1}g(u) \cos\left[\phi(\nu,u) t+\epsilon \pi u\right] du\nonumber\\
&\sim& \Re\left\{\int_0^{1}g(u)e^{i \epsilon \pi u} e^{i\phi(\nu,u) t} du\right\},
\label{h_cont} 
\end{eqnarray}
and expand $\phi(\nu,u)$ in the vicinity of $u_0$,
\begin{eqnarray*}
\phi(\nu,u)&\sim& \phi(\nu,u_0)+\frac{1}{2}\frac{\partial^2\phi(\nu,u_0)}{\partial u^2}(u-u_0)^2\\
&=&\phi_0(\nu)+\frac{1}{2}\phi''_0(\nu)(u-u_0)^2,
\end{eqnarray*}
being $u_0$ the point for which, given $\nu$, $\phi(\nu,u)$ has its maximum: 
\begin{eqnarray}
\frac{\partial\phi(\nu,u_0)}{\partial u}
&=& \pi (2\nu-1) +\frac{\pi \cos \pi u_0}{\sqrt{1+\cos^2 \pi u_0}}=0.
\label{phi_max}
\end{eqnarray}
From Eq.~\eqref{phi_max} we have
\begin{equation}
\cos \pi u_0 =\frac{1-2\nu}{2\sqrt{\nu(1-\nu)}},
\end{equation}
and
\begin{equation}
\sin \pi u_0 =\frac{1}{2}\sqrt{\frac{1-2(1-2\nu)^2}{\nu(1-\nu)}}.
\label{App_sin_u0}
\end{equation}
Equation~\eqref{App_sin_u0} tells us that the validity of the present approximation is restricted to values of $\nu$ for which one has $1-2(1-2\nu)^2> 0$, that is,
\begin{equation}
\frac{1}{2}\left(1-\frac{1}{\sqrt{2}}\right)<\nu<\frac{1}{2}\left(1+\frac{1}{\sqrt{2}}\right).
\end{equation}
Also from Eqs.~\eqref{omega_r} and~\eqref{App_sin_u0} we get
\begin{equation}
\sin \omega_0 =\frac{1}{2}\sqrt{\frac{1-2(1-2\nu)^2}{2\nu(1-\nu)}},
\end{equation}
as well as
\begin{equation}
\cos \omega_0 =\frac{1}{2\sqrt{2\nu(1-\nu)}},
\end{equation}
expressions that will be helpful in forthcoming derivations.
 
Now we can fully evaluate Eq.~\eqref{h_cont} under the above premises:
\begin{eqnarray}
h(n,t)&\sim& \Re\left\{\int_0^{1}g(u)e^{i \epsilon \pi u} e^{i\phi(\nu,u) t} du\right\} \nonumber\\
&\sim& \Re\left\{\int_0^{1}g(u_0)e^{i \epsilon \pi u_0} e^{i t\left[\phi_0(\nu)+\frac{1}{2}\phi''_0(\nu)(u-u_0)^2\right]} du\right\}\nonumber \\
&\sim& \Re\left\{g(u_0)e^{i \left[\epsilon \pi u_0+\phi_0(\nu) t\right]}\int_{-\infty}^{\infty} e^{\frac{i t}{2}\phi''_0(\nu)(u-u_0)^2} du\right\}\nonumber \\
&=&\sqrt{\frac{2 \pi}{t|\phi''_0(\nu)|}} g(u_0) \cos\left[\phi_0(\nu) t+\epsilon \pi u_0-\frac{\pi}{4}\right], 
\label{h_result}
\end{eqnarray}
with
\begin{eqnarray}
\phi''_0(\nu) &=&-4 \pi^2 \nu(1-\nu)\sqrt{1-2(1-2\nu)^2}.
\end{eqnarray}
The approximate versions of Eqs.~\eqref{Sol_Psi0} and~\eqref{Sol_Psi1} are
\begin{eqnarray}
\psi_{0}(n,t)&\sim&\frac{a}{\sqrt{t}} \sqrt{\frac{2(1-\nu)}{\pi \nu\sqrt{1-2(1-2\nu)^2}}} \cos\left[\phi_0(\nu) t-\pi/4\right]\nonumber \\
&+&\frac{b}{\sqrt{t}} \sqrt{\frac{2}{\pi \sqrt{1-2(1-2\nu)^2}}} \nonumber \\
&\times&\cos\left[\phi_0(\nu) t-\pi/4+\pi u_0\right], \label{Sol_App0}
\end{eqnarray}
and
\begin{eqnarray}
\psi_{1}(n,t)&\sim&\frac{a}{\sqrt{t}} \sqrt{\frac{2}{\pi \sqrt{1-2(1-2\nu)^2}}} \nonumber \\
&\times&\cos\left[\phi_0(\nu) t-\pi/4-\pi u_0\right]\nonumber \\
&+& \frac{b}{\sqrt{t}} \sqrt{\frac{2\nu}{\pi (1-\nu)\sqrt{1-2(1-2\nu)^2}}}\nonumber \\
&\times& \cos\left[\phi_0(\nu) t-\pi/4\right],
\label{Sol_App1}
\end{eqnarray}
and they follow from Eq.~\eqref{h_result} once one realizes that if 
\begin{equation*}
g(u)= \frac{1}{2-\sqrt{2} \cos(\omega_u -\pi u)},
\end{equation*}
one gets
\begin{equation*}
g(u_0)= 2(1-\nu);
\end{equation*}
if
\begin{equation*}
g(u)= \frac{\sqrt{2} \cos \omega_u -\cos \pi u}{2-\sqrt{2} \cos(\omega_u -\pi u)},
\end{equation*}
one has
\begin{equation*}
g(u_0)= 2\sqrt{\nu(1-\nu)};
\end{equation*}
and finally if
\begin{equation*}
g(u)= \frac{\left(\sqrt{2} \cos \omega_u -\cos \pi u\right)^2}{2-\sqrt{2} \cos(\omega_u -\pi u)},
\end{equation*}
one obtains
\begin{equation*}
g(u_0)= 2\nu.
\end{equation*}

The case in which $a=1/\sqrt{2}$, and $b=i/\sqrt{2}$ attracts much of our interest in the main text, so let us derive explicit approximate expressions for $\left|\psi_{0,1}(n,t)\right|^2$, and $\rho(n,t)=\left|\psi_0(n,t)\right|^2+\left|\psi_1(n,t)\right|^2$: 
\begin{eqnarray}
\left|\psi_0(n,t)\right|^2&\sim&\frac{1}{t} \frac{1}{\pi \nu\sqrt{1-2(1-2\nu)^2}} \nonumber \\
&\times&\left[(1-\nu)\cos^2 A+\nu\cos^2 (A+\pi u_0)\right],\nonumber \\
\label{Psi_0_App}
\end{eqnarray}
\begin{eqnarray}
\left|\psi_1(n,t)\right|^2&\sim&\frac{1}{t} \frac{1}{\pi (1-\nu)\sqrt{1-2(1-2\nu)^2}} \nonumber \\
&\times&\left[\nu\cos^2 A+(1-\nu)\cos^2 (A-\pi u_0)\right],\nonumber \\
\label{Psi_1_App}
\end{eqnarray}
and
\begin{eqnarray}
\rho(n,t)&\sim&\frac{1}{t} \frac{1}{\pi \nu(1-\nu)\sqrt{1-2(1-2\nu)^2}} \nonumber \\
&\times&\left\{\left[(1-\nu)^2+\nu^2\right]\cos^2 A\right.\nonumber \\
&+&\left.\nu(1-\nu)\left[\cos^2 (A-\pi u_0)+\cos^2 (A+\pi u_0)\right]\right\},\nonumber \\
\label{Prob_rough}
\end{eqnarray}
where we have introduced $A\equiv \phi_0(\nu) t-\pi/4$ to keep the expressions reasonably readable. 

The formula for $\rho(n,t)$ can be distilled even more. Let us expand the cosine terms in Eq.~\eqref{Prob_rough} 
\begin{eqnarray*}
&&\cos^2 (A-\pi u_0)+\cos^2 (A+\pi u_0)\\
&=&2 \cos^2 A \cos^2\pi u_0+2\sin^2 A \sin^2\pi u_0\\
&=&\frac{(1-2\nu)^2}{2\nu(1-\nu)}\cos^2 A+\frac{1-2(1-2\nu)^2}{2\nu(1-\nu)}\sin^2 A,
\end{eqnarray*}
to obtain in the first place
\begin{eqnarray}
\rho(n,t)&\sim&\frac{1}{t} \frac{1}{2\pi \nu(1-\nu)\sqrt{1-2(1-2\nu)^2}} \nonumber \\
&\times&\left\{\left[1+2(1-2\nu)^2\right]\cos^2 A\right.\nonumber \\
&+&\left.\left[1-2(1-2\nu)^2\right]\sin^2 A\right\},\nonumber \\
&=&\frac{1}{t} \frac{1}{2\pi \nu(1-\nu)\sqrt{1-2(1-2\nu)^2}} \nonumber \\
&\times&\left[1+2(1-2\nu)^2\cos 2A\right],\nonumber \\
\end{eqnarray}
but since
\begin{eqnarray*}
\cos 2A&=&\cos\left[ 2\phi_0(\nu) t-\pi/2\right]\\
&=&\sin\left[2\phi_0(\nu) t\right],
\end{eqnarray*}
we obtain
\begin{eqnarray}
\rho(n,t)\sim\bar{\rho}(n,t)&\equiv&\frac{1}{t} \frac{1}{2\pi \nu(1-\nu)\sqrt{1-2(1-2\nu)^2}} \nonumber \\
&\times&\left\{1+2(1-2\nu)^2\sin\left[2\phi_0(\nu) t\right]\right\}.\nonumber \\
\label{Prob_App}
\end{eqnarray}
Note how from $\bar{\rho}(n,t)$ we can define two new functions $\bar{\rho}_{\rm min}(n,t)$, and  $\bar{\rho}_{\rm max}(n,t)$,
\begin{eqnarray}
\bar{\rho}_{\rm max}(n,t)&\equiv&\frac{1}{t} \frac{1+2(1-2\nu)^2}{2\pi \nu(1-\nu)\sqrt{1-2(1-2\nu)^2}},\\
\label{Prob_App_Sup}
\bar{\rho}_{\rm min}(n,t) 
&\equiv&\frac{1}{t} \frac{\sqrt{1-2(1-2\nu)^2}}{2\pi \nu(1-\nu)}\\
\label{Prob_App_Inf}
&=&\frac{2}{\pi t} \sin 2\omega_0,
\label{Prob_App_Inf_Alt}
\end{eqnarray}
in such a way $\bar{\rho}_{\rm min}(n,t)\leq \bar{\rho}(n,t) \leq \bar{\rho}_{\rm max}(n,t)$.

Finally, let us derive Eq.~\eqref{Exit_Prob_App_Inf_Main}. This expression corresponds to a lower approximate bound of Eq.~\eqref{Psi_1_App}  when $\nu<1/2$. To this end we have to concentrate our attention in the case for which $\cos(A-\pi u_0)=0$,
\begin{eqnarray}
\left|\psi_1(n,t)\right|^2&\gtrsim&\frac{1}{t} \frac{\nu}{\pi (1-\nu)\sqrt{1-2(1-2\nu)^2}}\cos^2 A \nonumber \\
&=&\frac{1}{t} \frac{\nu}{\pi (1-\nu)\sqrt{1-2(1-2\nu)^2}}\sin^2 \pi u_0\nonumber \\
&=&\frac{1}{t} \frac{\sqrt{1-2(1-2\nu)^2}}{4 \pi (1-\nu)^2}.
\label{Exit_Prob_App_Inf_App}
\end{eqnarray}
Equation~\eqref{Exit_Prob_App_Inf_Main} follows after the replacement of $\nu$ by $n/t$.


\begin{thebibliography}{00}

\bibitem{ADZ93} \Journal{Y. Aharonov, L. Davidovich, and N. Zagury}{1993}{Quantum random walks}{Phys. Rev. A}{48}{1687}{1690} 

\bibitem{NK03} \Journal{N. Konno}{2003}{Quantum Random Walks in One Dimension}{Quantum Inf. Process.}{1}{345}{354} 

\bibitem{JK03} \Journal{J. Kempe}{2003}{Quantum random walks: An introductory overview}{Contemp. Phys.}{44}{307}{327}

\bibitem{FAJ04} \Journal{A. P. Flitney, D. Abbott, and N. F. Johnson}{2004}{Quantum walks with history dependence}{J. Phys. A: Math. Gen.}{37}{7581}{7591} 

\bibitem{VA12} \Journal{S. E. Venegas-Andraca}{2012}{Quantum walks: a comprehensive review}{Quantum Inf. Process.}{11}{1015}{1106} 

\bibitem{PS97} \Journal{P. W. Shor}{1997}{Polynomial-time algorithms for prime factorization and discrete logarithms on a quantum computer}{SIAM J. Comp.}{26}{1484}{1509}

\bibitem{FG98} \Journal{E. Farhi, and S. Gutmann}{1998}{Quantum computation and decision trees}{Phys. Rev. A}{58}{915}{928} 

\bibitem{MR95} \Book{R. Motwani and P. Raghavan}{1995}{Randomized algorithms}{CUP}{New York} %

\bibitem{CFG03} \Journal{A. Childs, E. Farhi, and S. Gutmann}{2003}{An Example of the Difference Between Quantum and Classical Random Walks}{Quantum Inf. Process.}{1}{35}{43} 

\bibitem{SKW03} \Journal{N. Shenvi, J. Kempe, and K. B. Whaley}{2003}{Quantum random-walk search algorithm}{Phys. Rev. A}{67}{052307}{} 

\bibitem{AMB10} \Journal{E. Agliari, A. Blumen, and O. N\"ulken}{2010}{Quantum-walk approach to searching on fractal structures}{Phys. Rev. A}{82}{012305}{} 

\bibitem{MNRS11} \Journal{F. Magniez, A. Nayak, J. Roland, and M. Santha}{2011}{Search via quantum walk}{SIAM J. Comp.}{40}{142}{164}

\bibitem{A05} D. Applebaum, in {\it  L\'evy Processes in Euclidean Spaces and Groups\/}, Lect. Notes Math. {\bf 1865},  edited by U. Franz and N. Sch\"urmann (Springer, Berlin, 2005), pp. 1-98.

\bibitem{BCGJW04} \Journal{E. Bach, S. Coppersmith, N. P. Goldschen, R. Joynt, and J. Watrous}{2004}{One-dimensional quantum walks with absorbing boundaries}{J. Comput. Syst. Sci.}{69}{562}{592} 

\bibitem{KB06} \Journal{H. Krovi and T. A. Brun}{2006}{Hitting time for quantum walks on the hypercube}{Phys. Rev. A}{73}{032341}{} 


\bibitem{KS11} \Journal{L. C. Kwek and Setiawan}{2011}{One-dimensional quantum walk with a moving boundary}{Phys. Rev. A}{84}{032319}{} 

\bibitem{CH46} \Book{R. Courant and D. Hilbert}{1953}{Methods of mathematical physics}{CUP}{Cambridge}
\end{thebibliography}
\end{document}